\documentclass[twocolumn,aps,pr,superscriptaddress,preprintnumbers,nofootinbib,10pt]{revtex4-2}
\usepackage{multirow}
\usepackage{amsmath}
\usepackage{amssymb}
\usepackage[dvipdf,dvips]{graphicx}
\usepackage{color}
\usepackage{hyperref}
\usepackage{url}
\usepackage{slashed}
\usepackage{subfigure}
\usepackage[usenames,dvipsnames]{xcolor}
\usepackage{amsmath}
\usepackage{amsfonts}
\usepackage{float} 
\usepackage{amssymb}
\usepackage{epsfig}
\usepackage{graphics}
\usepackage{euscript}
\usepackage{slashed}
\usepackage{epstopdf}
\usepackage[utf8]{inputenc}
\allowdisplaybreaks
\usepackage[normalem]{ulem}
\usepackage{pifont}
\usepackage{dsfont}
\usepackage{MnSymbol}
\usepackage{verbatim}
\usepackage{graphicx}
\usepackage{latexsym}
\usepackage{tikz-feynman}

\begin{document}

\title{Four-gluon vertex from the Curci-Ferrari model at one-loop order} 

	\author{Nahuel Barrios}  \affiliation{Instituto de Física, Facultad de Ingeniería, Universidad de la República, J. H. y Reissig 565, 11000 Montevideo, Uruguay} \affiliation{UERJ $–$ Universidade do Estado do Rio de Janeiro, Rua São Francisco Xavier 524, 20550-013, Rio de Janeiro, Brazil}
	
	\author{Philipe De Fabritiis}  \affiliation{UERJ $–$ Universidade do Estado do Rio de Janeiro, Rua São Francisco Xavier 524, 20550-013, Rio de Janeiro, Brazil} \affiliation{CBPF $-$ Centro Brasileiro de Pesquisas Físicas, Rua Dr. Xavier Sigaud 150, 22290-180, Rio de Janeiro, Brazil}
	
	\author{Marcela Peláez} \affiliation{Instituto de Física, Facultad de Ingeniería, Universidad de la República, J. H. y Reissig 565, 11000 Montevideo, Uruguay}

	\begin{abstract}

 We compute the four-gluon vertex from the Curci-Ferrari model at one-loop order for a collinear configuration. Our results display a good agreement with the first lattice data for this vertex, released very recently (arXiv:2401.12008). A noteworthy novelty of our work is that we can provide analytical expressions for the four-gluon vertex in collinear configurations, together with a renormalization scheme that allows us to perform reliable perturbative computations even in the infrared regime. We observe an infrared suppression in the form factor associated with the tree-level four-gluon tensor with a possible zero-crossing in the deep infrared which demands new lattice investigations to be confirmed. Moreover, we report an infrared divergence in the completely symmetric tensor form factor due to the ghost-loop contributions. These results come as predictions since previous two-point correlations fix all the available parameters of the model, up to an overall constant factor. 
 
 \end{abstract}

	\maketitle	
	
	\section{Introduction}\label{SecIntro}

The infrared regime (IR) of Quantum Chromodynamics (QCD) poses one of the most challenging open problems of Physics, namely, understanding the mechanism behind color confinement. In addition to this paramount question, other phenomena such as the spontaneous breaking of chiral symmetry and the emergence of fundamental mass scales also present major challenges. The use of different approaches can shed light on distinct aspects of these problems, and their synergy can help us build a clearer understanding of such mysteries.

It is well-known that the Faddeev-Popov (FP) procedure~\cite{Faddeev67}, although extremely successful in high energies thanks to asymptotic freedom~\cite{Politzer73, Gross73}, cannot describe the IR regime of Yang-Mills (YM) theories. In particular, lattice simulations established two remarkable facts that the standard FP approach cannot explain: i) it is possible to define a gauge coupling that remains finite and moderate at all scales; ii) the gluon propagator at vanishing momentum has a finite nonzero value~\cite{Bogolubsky09}. These striking facts suggest that perturbation theory can be used to extract information from the IR regime and that the gluon behaves as if it develops a screening mass in the IR.

The Curci-Ferrari (CF) model in the Landau gauge~\cite{Tissier10, Tissier11, Curci76} is
the simplest renormalizable deformation of the FP Lagrangian that allows for a description of such remarkable features, and whose results have been successfully confirmed by lattice data~\cite{Marcela21}.
The main advantage of the CF model is to provide a reliable and controllable perturbative approach to investigate the IR regime of YM theories. For recent achievements, see Refs.~\cite{Marcela23, Marcela21, Marcela21b, Marcela17, Reinosa17}.
 
A reasonable way to extract information from the IR regime of YM theories is to compute its correlation functions. In the last decades, much progress has been made using many approaches, such as Lattice QCD~\cite{Cucchieri07,  Cucchieri08a, Cucchieri08b, Cucchieri08c, Oliveira11, Oliveira12, Duarte16a}, the CF model~\cite{Tissier10, Tissier11, Curci76, Marcela21}, Schwinger-Dyson
equations~\cite{Roberts94, Alkofer01, Fischer06, Binosi09, Huber20b, Ferreira23}, the functional renormalization group~\cite{Berges02, Pawlowski07, Braun10, Gies12, Dupuis20}, the screened mass expansion~\cite{Siringo16a,Siringo16b,Comitini21} and the refined Gribov-Zwanziger model~\cite{Dudal08, Dudal11, Vandersickel12, Capri15, Capri16}. Nowadays, it can be safely said that we have achieved a sensible understanding of the two-point correlation functions~\cite{Barrios21, Gracey19, Marcela14, Cyrol18, Aguilar08, Fischer09,Aguilar15, Oliveira19, Philipe23, Comitini24}. Progress has also been made in computing higher correlation functions such as the three-gluon and the ghost-gluon vertices~\cite{Barrios22, Figueroa22, Barrios20, Barrios24, Mintz18, Marcela13, Blum14, Schleifenbaum05, Cucchieri06, Athenodorou16, Duarte16b, Binosi13,Aguilar19a, Aguilar19, Aguilar21, Aguilar23, Maas2022}.

However, the four-point gluon correlation is still poorly understood, mainly due to the proliferation of allowed tensorial structures in this case (more than one hundred for general kinematics~\cite{Gracey14}). In fact, there are only a few semi-analytical works on the subject~\cite{Aguilar24, Binosi14, Cyrol15, Cyrol16, Huber20, Kellermann08}, and the first lattice data for this vertex came up just very recently~\cite{Orlando24} (for preliminary analysis, see Refs.~\cite{Orlando23, Catumba21}).
 
The main goal here is to compute the four-gluon vertex at one-loop order within the CF model, describing its main features and comparing our results with the first lattice data for this quantity~\cite{Orlando24}. We find a good agreement with lattice data, with a relatively low computational cost, attesting to the efficiency of the CF model in describing the IR regime of YM theories through perturbation theory. Our results are predictions of the CF model since the previous evaluation of the two-point correlation functions fixes all the parameters of the model, up to an overall constant factor. We observe an IR suppression of the tree-level four-gluon tensor form factor, with an apparent zero-crossing in the deep IR. Moreover, we report an IR divergence in the totally symmetric tensor form factor due to the ghost-loop contributions. Finally, we show that we can provide analytical expressions for any collinear configuration.

This work is organized as follows. In Sec.~\ref{SecTheory}, we present the CF model in the Landau gauge and general aspects of the four-gluon vertex. Sec.~\ref{SecRenorm} addresses the renormalization and the renormalization group improvement. We present our results in Sec.~\ref{SecResults}, comparing them with fresh lattice data~\cite{Orlando24}. In Sec.~\ref{SecConclusions}, we state our conclusions. 
In App.~\ref{AppAnalytical}, we show that we can provide analytical expressions at one-loop for any collinear configuration.

\section{The Curci-Ferrari model and the four-gluon vertex}\label{SecTheory}

Let us consider the Curci-Ferrari Lagrangian in the Landau gauge with Euclidean metric:
	\begin{align}\label{CFlag}
		\mathcal{L} \!=\! \frac{1}{4} F_{\mu \nu}^a F_{\mu \nu}^a + i h^a \partial_\mu A_\mu^a + \partial_\mu \bar{c}^a (D_\mu c)^a + \frac{1}{2} m^2 A_\mu^a A_\mu^a,
	\end{align}
where the covariant derivative in the adjoint representation is given by $D_\mu^{ab} = \delta^{ab} \partial_\mu - g f^{abc} A_\mu^c$, the structure constants for the $SU(N)$ group are denoted by $f^{abc}$, and the bare coupling and the gluon mass are given by $g$ and $m$, respectively. The field strength is given by $F_{\mu\nu}^a = \partial_\mu A_\nu^a - \partial_\nu A_\mu^a + g f^{abc} A_\mu^b A_\nu^c$, the FP ghosts are $c, \bar{c}$, and $h^a$ is the Nakanishi-Lautrup field enforcing the Landau gauge condition $\partial_\mu A_\mu^a= 0$. We remark that the CF model is renormalizable in 4 dimensions, it recovers the usual FP theory in the UV regime and its mass term enables the use of reliable perturbation theory at all scales.

We are interested in computing the four-gluon one-particle irreducible (1PI) Green function at one-loop and comparing it with the very recent lattice data~\cite{Orlando24}. However, in the numerical simulations, only the full Green functions can be accessed. To obtain information concerning the 1PI Green function from the full Green function without extra assumptions, the computation must be restricted to suitable momentum configurations~\cite{Orlando24}.

The full Green function includes a connected part plus disconnected contributions. Adopting a kinematical configuration such that $p_i + p_j \neq 0, \forall i,j$ where $p_i$ is the momentum associated with the $i$-th external leg, the contributions due to the disconnected part can be discarded. Now, the connected part itself can be written as a sum of a contribution with the four-gluon 1PI Green function plus contributions containing the three-gluon 1PI Green function. In the Landau gauge, due to the orthogonality of the gluon propagator, considering that all external momenta are proportional, the contributions including the three-gluon vertex can be safely discarded, allowing to access the four-gluon 1PI Green function~\cite{Orlando24}.

In this work, for simplicity, we adopt the collinear configuration $\left(p_1, p_2, p_3, p_4\right) = \left(p, p, p, -3p\right)$ with gauge group $SU(3)$ in four space-time dimensions. Such configuration already allows us to state the four-gluon vertex main features and compare with lattice data. However, we stress that any collinear configuration satisfying the above restrictions could be chosen. Due to the orthogonality of the Landau gauge gluon propagator and the simple momentum configuration adopted, the only non-vanishing contributions of interest to compare with lattice simulations will be the ones proportional to the product of metrics in the Lorentz sector. Although there are still many possible tensorial structures due to the color sector, for the configuration $(p,p,p,-3p)$ with $SU(3)$ at one-loop, only two of them will appear in our computation~\cite{Pascual80, Binosi14}.

To compute the four-gluon 1PI Green function in perturbation theory at one-loop order, one needs to calculate 18 Feynman diagrams\footnote{We evaluated the diagrams using our own code in {\it Mathematica}  and independently by using the package {\it FeynCalc}~\cite{FeynCalc1, FeynCalc2, FeynCalc3}.} (see~\cite{Binosi14} for the relevant diagrams). 
For the $(p,p,p,-3p)$ configuration with gauge group $SU(3)$, the only tensorial structures appearing at one-loop accuracy are the tree-level four-gluon vertex $H_{\mu \nu \rho \sigma}^{abcd}$, and the totally symmetric tensor $G_{\mu \nu \rho \sigma}^{abcd}$, defined respectively as,
\begin{align}\label{Htensor}
	 	H^{a b c d}_{\mu \nu \rho \sigma} &= f^{adx} f^{cbx} \left(\delta_{\mu \rho} \delta_{\nu \sigma} - \delta_{\mu \nu} \delta_{\rho \sigma} \right) \nonumber \\
	 	&+  f^{abx} f^{dcx} \left(\delta_{\mu \sigma} \delta_{\nu \rho} - \delta_{\mu \rho} \delta_{\nu \sigma}  \right) \nonumber \\
	 	&+  f^{acx} f^{dbx} \left(\delta_{\mu \sigma} \delta_{\nu \rho} - \delta_{\mu \nu} \delta_{\rho \sigma} \right).
	 \end{align}
 and
   \begin{align} \label{Gtensor}
	 	G^{a b c d}_{\mu \nu \rho \sigma} &= \left(\delta^{ab} \delta^{cd} + \delta^{ac} \delta^{bd} + \delta^{ad} \delta^{bc}\right) R_{\mu \nu \rho \sigma},
	 \end{align}
  where we have $ R_{\mu \nu \rho \sigma} = \left(\delta_{\mu \nu} \delta_{\rho \sigma} + \delta_{\mu \rho} \delta_{\nu \sigma} + \delta_{\mu \sigma } \delta_{\nu \rho}\right)$. Therefore, the four-gluon 1PI Green function in this specific configuration can be written as
\begin{align}\label{4gluon3pBare}
    \Gamma_{\mu \nu \rho \sigma}^{abcd} (p,p,p,-3p) = H(p) H_{\mu \nu \rho \sigma}^{abcd} + G(p) G_{\mu \nu \rho \sigma}^{abcd}.
\end{align}
It should be remarked that these tensors are orthogonal.

\section{Renormalization and Renormalization Group} \label{SecRenorm}

Let us define the renormalized quantities in terms of the bare ones as,
\begin{align}
    A_B^{\mu,a} = \sqrt{Z_A} A_R^{\mu,a}, \quad c_B^a = \sqrt{Z_c} c_R^a, \quad \bar{c}_B^a = \sqrt{Z_c} \bar{c}_R^a.
\end{align}
Regarding the coupling and the mass, we can write,
\begin{align}
    g_B = Z_g g_R, \quad m_B^2 = Z_{m^2} m_R^2.
\end{align}
For the renormalized 1PI four-point Green function we have,
\begin{align}\label{gammaRtoB}
    \left(\Gamma^{R}\right)^{abcd}_{\mu \nu \rho \sigma} (p,\mu)= Z_A^2 \left(\Gamma^{B}\right)^{abcd}_{\mu \nu \rho \sigma}(p),
\end{align}
where the dependence on the renormalization scale $\mu$ is introduced by means of the $Z$-factors.

A distinguishing feature of the CF model is that it allows for a renormalization scheme in which the renormalization flow is regular at all scales~\cite{Tissier11,Reinosa17}. In the infrared safe (IRS) renormalization scheme we have the following renormalization conditions
\begin{align}\label{IRS1}
    D_{\rm A}^{-1}(p=\mu) = \mu^2 + m^2(\mu), \quad D_{\rm c}^{-1}(p=\mu) = \mu^2,
\end{align}
where $D_{\rm A}$ and $D_{\rm c}$ are the gluon and ghost propagators, as well as the non-renormalization conditions:
\begin{align}\label{IRS2}
    Z_g \sqrt{Z_A} Z_c = 1,  \quad
    Z_{m^2} Z_A Z_c = 1.
\end{align}
Notice that the above constraints are valid not only for the divergent part (as the usual non-renormalization condition), but also for the finite parts in this scheme.


All the UV divergences are proportional to the tensor $H_{\mu \nu \rho \sigma}^{abcd}$, as expected by renormalizability arguments since this is the only tensor that appears at tree-level. The bare form factor associated with this tensorial structure at one-loop order can be written as
\begin{align}
H(p) =g_B^2\left( 1 + \frac{g_B^2}{\varepsilon} H_{div} + g_B^2 H_{fin}(p)\right),
\end{align}
where $g_B^4 H_{div}$ and $g_B^2(1 + g_B^2) H_{fin}(p)$ correspond to the divergent and finite parts of $H(p)$, respectively. Consequently, the 1PI Green function~\eqref{gammaRtoB} at one-loop reads
\begin{align}\label{eq:Gamma_bare_oneloop}
    \left(\Gamma^{B}\right)^{abcd}_{\mu \nu \rho \sigma} = &g^2_B\left(1  + \frac{g^2_B}{\varepsilon} H_{div} + g^2_B H_{fin}\right) H_{\mu \nu \rho \sigma}^{abcd} \nonumber \\
    &+ g_B^4 G_{fin} G_{\mu \nu \rho \sigma}^{abcd}.
\end{align}

At one-loop order, the renormalization factors can be written in the following way
\begin{align}\label{eq:Zs_oneloop}
    Z_A = 1 + g_R^2 \, \delta Z_A, \quad Z_g = 1 + g_R^2 \, \delta Z_g.
\end{align}

By inserting the relations \eqref{eq:Zs_oneloop} and \eqref{eq:Gamma_bare_oneloop} into Eq. \eqref{gammaRtoB} and retaining terms up to order $g^4_R$ we obtain
\begin{align}
    & \left(\Gamma^{R}\right)^{abcd}_{\mu \nu \rho \sigma}(p,\mu) \!=    g^4_R G_{fin} G_{\mu \nu \rho \sigma}^{abcd} \nonumber \\
    & + \left[g^2_R + g^4_R \left(2 \delta Z_A + 2 \delta Z_g + \frac{H_{div}}{\varepsilon} + H_{fin} \right)\right] H_{\mu \nu \rho \sigma}^{abcd},
\end{align}
where we have made explicit the dependence of the renormalized correlation function on the momentum $p$ and the renormalization scale $\mu$. The quantities $\delta Z_A$ and $\delta Z_c$ are fixed by the renormalization of the two-point functions at one-loop order in the IRS scheme \cite{Tissier11}.
Needless to say, the divergent parts perfectly cancel in the above expression. One can immediately obtain $Z_g$ and $Z_{m^2}$ resorting to the non-renormalization conditions~\eqref{IRS2} of the IRS renormalization scheme.

To avoid large logarithms of the form $\log \left(\frac{p^2}{\mu^2}\right)$ in the UV, which could potentially spoil the use of the perturbative framework, we must adopt a renormalization scale of the type $\mu \approx p$. This choice, nonetheless, is not suitable for comparisons with lattice simulations, where a fixed renormalization scale $\mu_0$ is required. This obstacle can be overcome by employing the Callan-Symanzik equation~\cite{Callan70, Symanzik70}, which in the case of a purely gluonic vertex function with $n_A$ external legs reads
\begin{equation}
    \left( \mu \partial_\mu -\frac{1}{2}n_A \gamma_A+\beta_g \partial_g+\beta_{m^2}\partial_{m^2}\right)\Gamma^{(n_A)}=0,
\end{equation}
where we omitted the sub-index ``R" to refer to renormalized quantities, as we will do from now on. We introduced the usual $\beta$-functions and the anomalous dimension $\gamma_A$:
\\
\begin{align}
    \beta_X(g,m^2) &\equiv \mu \frac{dX}{d\mu}\bigg|_{g_B,m^2_B}, \\
    \gamma_A(g,m^2) &\equiv \mu \frac{d \log Z_A}{d\mu}\bigg|_{g_B,m^2_B},
\end{align}
with $X \in \{g,m^2\}$. The solution for the 1PI four-gluon correlation function is
\begin{align}\label{eq:sol_CS}
    \Gamma_{\mu \nu \rho \sigma}^{abcd}&(p,\mu_0,g_0,m^2_0)=\nonumber\\
    &z_A(\mu,\mu_0)^{-2}\Gamma_{\mu \nu \rho \sigma}^{abcd}(p,\mu,g(\mu),m^2(\mu))),
\end{align}
where
\begin{equation}
    z_A(\mu) = \exp \left[\int_{\mu_0}^{\mu} \frac{d\mu'}{\mu'} \, \gamma_A \left( g(\mu'), m^2(\mu') \right)\right].
\end{equation}
The advantage of the relation \eqref{eq:sol_CS} is that it enables us to safely evaluate $\Gamma_{\mu \nu \rho \sigma}^{abcd}(p,\mu_0,g_0,m^2_0)$ in perturbation theory even for scales such that $p\gg\mu_0.$

By using the conditions \eqref{IRS2} and the independence of the bare quantities with the scale $\mu$, one can prove that
\begin{equation}
\gamma_A=2\left(\frac{\beta_{m^2}}{m^2}-\frac{\beta_g}{g}\right),    
\end{equation}
from which
\begin{equation}
    z_A(\mu,\mu_0)=\frac{m^4(\mu)}{m^4(\mu_0)}\frac{g^2(\mu_0)}{g^2(\mu)}.
\end{equation}
As a result, we can write
\begin{align}
    \Gamma_{\mu \nu \rho \sigma}^{abcd}&(p,\mu_0,g_0,m^2_0)=\nonumber\\
    &\frac{m^8(\mu_0)}{m^8(\mu)}\frac{g^4(\mu)}{g^4(\mu_0)}\Gamma_{\mu \nu \rho \sigma}^{abcd}(p,\mu,g(\mu),m^2(\mu))).
\end{align}

The parameters $g(\mu)$ and $m^2(\mu)$ can be obtained by integrating the beta functions with initial conditions $g(\mu_0)$ and $m^2(\mu_0)$, defined at the scale $\mu_0$. To fix such conditions, we use a one-loop fitting for the lattice data of the gluon and ghost two-point functions, as was done in \cite{Tissier11}. We proceed in this manner since lattice data for these correlations is far more precise than for higher correlation functions, allowing for an accurate fitting. Furthermore, this allows us to make a prediction for the four-gluon vertex, up to an overall normalization constant, instead of merely fitting the lattice data, putting the theory to a more precise and sharp test. In this work we opt for the prescription $\mu=\sqrt{p^2+m^2(\mu_0)}$ which fulfils the condition $\mu \approx p $ as $p^2\gg m^2(\mu_0) $ but also avoids potential large logarithms of the type $\log\left(\frac{m^2}{\mu^2} \right)$ in the IR, as explained in~\cite{Barrios22}. The aforementioned fits lead to the values $g(1 {\rm GeV}) = 3.97$ and $m^2(1 {\rm GeV}) = 0.35^2 \, {\rm GeV}^2$, where we set $\mu_0=1$ GeV.

\newpage
 
\section{Results and discussion}\label{SecResults}

The main goal of this work is to explicitly compute the four-gluon 1PI Green function~\eqref{4gluon3pBare} at one-loop order in the CF model for the configuration $(p, p, p, -3p)$ with gauge group $SU(3)$ in four space-time dimensions and compare it with the brand new lattice data that was released recently~\cite{Orlando24}. Our results come as a prediction of the CF model since all its parameters were previously fixed by two-point correlators, up to an overall normalization constant. 
This global factor is needed to compare our results obtained in the IRS renormalization scheme with the lattice data obtained using a different scheme. This factor was determined by minimizing a joint error $\chi^2=\frac 1 2 (\chi_H^2+\chi_G^2)$, where $\chi_H^2$ and $\chi_G^2$ measure the error between the lattice data and the CF outcomes. We found the overall constant for $H$ to be $\mathcal{N}=0.10695$, which leads to a factor for $G$ of $\frac{10800}{3888} \mathcal{N}$, see \cite{Orlando24}.

\begin{figure}[t!]
 	\begin{minipage}[b]{1.0\linewidth}
 		\includegraphics[width=\textwidth]{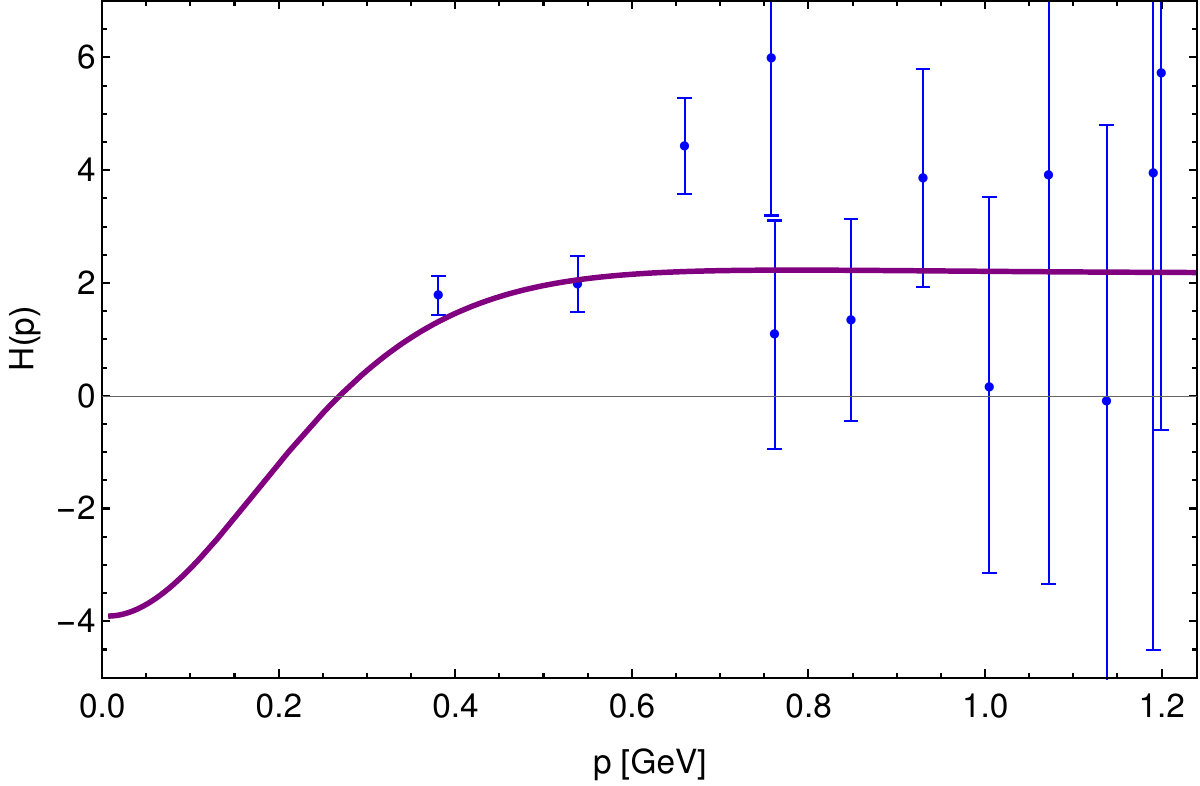}
 	\end{minipage} \hfill
 	\caption{Form factor $H(p)$ as predicted within the CF model at one-loop, compared with the $SU(3)$ lattice data of~\cite{Orlando24}.}
 	\label{Hgraph}
 \end{figure}

\begin{figure}[t!]
 	\begin{minipage}[b]{1.0\linewidth}
 		\includegraphics[width=\textwidth]{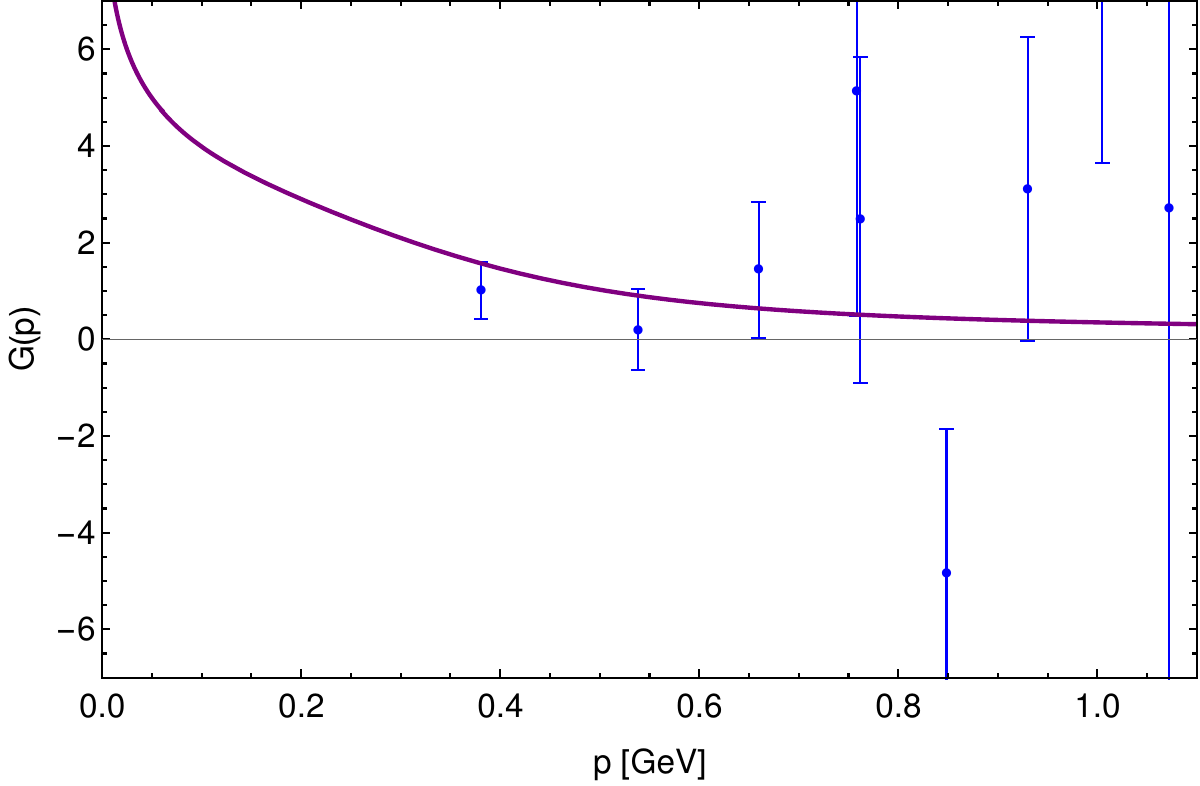}
 	\end{minipage} \hfill
 	\caption{Form factor $G(p)$ as predicted within the CF model at one-loop, compared with the $SU(3)$ lattice data of~\cite{Orlando24}.}
 	\label{Ggraph}
 \end{figure}

The four-gluon 1PI Green function in the configuration $(p,p,p,-3p)$ in $SU(3)$ can be written in terms of analytical expressions for the bare form factors $H(p)$ and $G(p)$, which can be found in the supplementary material~\cite{Supplement1}. The final results for the renormalized form factors $H(p)$ and $G(p)$ taking into account the renormalization group improvement can be found in Figs.~\ref{Hgraph} and~\ref{Ggraph}, along with their corresponding lattice data~\cite{Orlando24}. 
We show in App.~\ref{AppAnalytical} that our approach can provide analytical expressions for the four-gluon vertex at one-loop for any collinear configuration, showing that this remarkable feature is not a peculiarity of the simple configuration chosen here.
 

Let us discuss the main characteristics of our findings. Two interesting salient features show up in the IR. First, the form factor $H(p)$ is suppressed in the IR in comparison with its tree-level value, achieving a finite value at $p=0$, with an apparent zero-crossing in the deep IR which demands further investigations to be confirmed.
Second, there is an IR singularity in the form factor $G(p)$.
We remark that the lattice data, although being a pioneering work, has still a significant error, and only a few points below $1 {\rm GeV}$, where these peculiar aspects are present. These two features are also present in the three-gluon correlation~\cite{Barrios22, Figueroa22, Marcela13}, although in that case, both happen in the tree-level tensor form factor.

The IR suppression of the form factor $H(p)$ in comparison with its tree-level value is suggested by the lattice data~\cite{Orlando24} and was also reported in~\cite{Aguilar24}. The apparent zero-crossing that we observe in the deep IR was not reported in~\cite{Aguilar24, Binosi14, Cyrol15} and the present lattice data does not allow for a conclusion in this matter. However, this peculiar feature should be taken with a grain of salt because it could be an artifact of the approximations adopted here. A zero-crossing is also observed in the three-gluon correlation~\cite{Barrios22, Figueroa22, Marcela13} but in that case, it comes from the ghost-loop contribution to the tree-level structure.
This is different from the present case, where the change of sign is not due to ghost-loop contributions, which can only contribute to $G(p)$, as we will see. On the other hand, by increasing the momentum, $H(p)$ approaches the tree-level value, in agreement with~\cite{Aguilar24, Binosi14, Cyrol15}, as expected. 

The second feature, namely, the IR divergence in $G(p)$, is due to the masslessness of the ghosts that appear in the ghost-loop graphs, giving a contribution proportional to the totally symmetric tensor $G_{\mu \nu \rho \sigma}^{abcd}$ that blows up as we go in the limit $p \rightarrow 0$. 
This feature was also reported in Refs.~\cite{Aguilar24, Binosi14}. However, we did not observe a zero-crossing in this form factor as reported in these works, and no change of sign is suggested by the lattice data~\cite{Orlando24}. An IR divergence due to ghost-loop contributions is a feature also present in the three-gluon correlation~\cite{Barrios22, Figueroa22, Marcela13}. In the deep IR, the $G(p)$ behavior due to the ghost-loop contribution is given by:  
$G(p \rightarrow 0) \approx \frac{3 g^4}{256 \pi^2} \log\left(\frac{m^2}{p^2}\right)$.

Interestingly enough, when considering tensorial structures accessible to lattice simulations, the above-mentioned IR divergence can only affect the form factor associated with the totally symmetric tensor $G_{\mu \nu \rho \sigma}^{abcd}$ at one-loop order, not only for the simple configuration analyzed here but also for any collinear configuration. In fact, let us consider the ghost-loop contributions for the four-gluon vertex at one-loop. By the gluon-ghost vertex structure, there will be structure constants and an incoming momentum in each vertex. Now, for collinear configurations $\left( \alpha p, \beta p, \gamma p, -(\alpha + \beta + \gamma) p \right)$, contributions proportional to the four-vector $p$ will appear but should be ignored due to the orthogonality of the Landau gauge gluon propagator of the external legs since we aim to compare with lattice simulations, which can only access the full Green function.
Thus, we will obtain contributions that are proportional to $q_\mu q_\nu q_\rho q_\sigma$, where $q$ is the momentum running inside the loop. This is totally symmetric in the Lorentz indices and thus proportional to $ R_{\mu \nu \rho \sigma}$. Being totally symmetric in the Lorentz indices, the Bose symmetry of the external legs constrains the color structure to be also totally symmetric. Notice that the color structure cannot depend on the specific momentum configuration chosen. Thus, we can take the limit $p \rightarrow 0$ for simplicity, and ascertain that the color structure is of the form $\left( \delta^{ab} \delta^{cd} + \delta^{ac} \delta^{bd} + \delta^{ad} \delta^{bc} \right)$. Therefore, the case $p \neq 0$ will have the same color structure and we can say that the ghost-loop contributions (and its associated IR divergence) will be proportional to the totally symmetric tensor $G_{\mu \nu \rho \sigma}^{abcd}$ for any collinear configuration.

Finally, we close this section by briefly highlighting some of the previous findings regarding the four-gluon vertex in the literature. First of all, the recent pioneering work~\cite{Orlando24} provides lattice data for the four-gluon vertex in three collinear configurations (including the one studied here) for some form factors, reporting a dominance of the form factor $H(p)$ that is essentially constant in the range accessed, suggesting an IR divergence in $G(p)$, but not suggesting zero-crossings for $H(p)$ or $G(p)$. The form factor $F^{(1)}$ in our computation is a mere linear combination of the already presented form factors, with an associated plot of the same quality of Figs.~\ref{Hgraph} and~\ref{Ggraph}.

In Ref.~\cite{Binosi14}, the authors adopted the same collinear configuration we adopted here, and our analytical results for the Feynman graphs are compatible with the ones they report. However, they observed a large peak in the IR for $H(p)$ that we did not observe (nor the authors of~\cite{Aguilar24}), and it is not suggested by lattice data. 
In Ref.~\cite{Cyrol15}, the authors focused on the tree-level tensor form factor, discussed more general kinematical configurations, and provided a comparison between decoupling and scaling behaviors for the propagators. They obtained an IR suppression of $H(p)$ like us, but they reported a prominent peak in the IR which we did not observe. For the scaling solutions, they observed IR divergences in $H(p)$ that were not observed by us nor by~\cite{Aguilar24}. The authors of Refs.~\cite{Cyrol16, Huber20} report results in qualitative agreement with the ones presented in~\cite{Cyrol15}.

The very recent work~\cite{Aguilar24} focused on the class of collinear configurations using lattice data as input, providing a thorough discussion about the possible tensorial structures appearing in their setup. Restricting their attention to a subset of three form factors, the authors did a robust investigation on the emergence of planar degeneracy in the IR and discussed the construction of an effective charge. They reported an IR suppression of $H(p)$ and an IR divergence in $G(p)$ like us, but they observed a zero-crossing in $G(p)$ instead of in $H(p)$ as we reported here. They directly compared their results with Refs.~\cite{Binosi14, Cyrol15}, discussing the absence of the prominent peak that these authors found in the IR for $H(p)$, concluding that this difference could be due to the different approximations they adopted for the three-gluon vertex.

\vspace{-0.2cm}

\section{Conclusions}\label{SecConclusions}

We computed the four-gluon 1PI Green function in the Landau gauge at one-loop order using the Curci-Ferrari model for the kinematical configuration $(p,p,p,-3p)$ with gauge group $SU(3)$ in four dimensions and compared our results with the first lattice data for this vertex~\cite{Orlando24}.


Analytical expressions for the form factors $H(p)$ and $G(p)$ were provided. After performing the renormalization group improvement adopting the IRS renormalization scheme, we compared our results with the lattice data of Ref.~\cite{Orlando24}, achieving a good agreement. These results come as predictions of our model, up to an overall constant factor, since all the available parameters were fixed beforehand. Our results exhibit an IR suppression in the form factor $H(p)$ and an IR divergence in $G(p)$. These features were also reported by other semi-analytical works~\cite{Binosi14, Aguilar24}, and are also present in the three-gluon vertex, although they affect the same form factor in that case. Furthermore, we observed an apparent zero-crossing in the form factor $H(p)$ taking place in the deep IR that was not reported elsewhere, but this prediction should be taken with a grain of salt since it could be an artifact of the one-loop approximation or due to the particular renormalization scheme employed here. It would be enlightening to have new lattice data accessing this vertex in the deep IR to confirm these features.

Within our approach, we can provide analytical expressions at one-loop order for the four-gluon vertex for any collinear configuration $\left( \alpha p, \beta p, \gamma p, -(\alpha + \beta + \gamma)p \right)$ and also that there will be IR divergences in the form factor associated with the totally symmetric tensor $G_{\mu \nu \rho \sigma}^{abcd}$ for any collinear configuration at one-loop accuracy due to the ghost-loop graphs. Furthermore, although we focused on the particular configuration $\left(p, p, p, -3p\right)$ for simplicity, the same analysis could be done with any collinear configuration, and preliminary results indicate that they exhibit quite similar features.

This work comes as a first result for the four-gluon vertex within the CF model.
The pioneering work~\cite{Orlando24} opens the doors to more throughout lattice investigations, and we hope that soon there will be more precise data available for the IR regime. We plan to exhibit analytical results for other collinear configurations and investigate the possible emergence of planar degeneracy. We hope to report on these subjects soon and provide a more in-depth analysis in a forthcoming paper.

\begin{acknowledgments}
    The authors would like to thank O. Oliveira for kindly sharing the lattice data, and M. Tissier, N. Wschebor, and U. Reinosa for many interesting discussions. PDF is grateful for the warm hospitality and the financial support of the Universidad de la República (UdelaR), where part of this work was developed, to G. P. de Brito for helpful discussions, and to D. Dudal for valuable comments on the manuscript. The authors are grateful to the referee for the valuable report. This work has been partially supported by PEDECIBA and the ANII-FCE-166479 project. PDF acknowledges FAPERJ for financial support under the contract SEI-260003/000133/2024.
\end{acknowledgments}

\appendix

\section{Analytic expression}\label{AppAnalytical}

In general, it is expected that the four-gluon vertex in arbitrary kinematics cannot be expressed analytically, even at one-loop order. However, for the case of collinear configurations, analytical expressions can be given in the CF model, because all master integrals can be expressed in terms of the integrals $\mathcal{A}$ and $\mathcal{B}$ defined by:
\begin{align}
    \mathcal{A}(m^2)&= \!\! \int_q \!\!\ \frac{1}{q^2+m^2},\\
    \mathcal{B}(p,m_1^2,m_2^2)&= \!\! \int_q \!\!\ \frac{1}{(q^2+m_1^2)\left((q+p)^2+m_2^2\right)},
\end{align}
of which their analytical expression is known. We define 
\begin{equation}
    \int_q \equiv \Lambda^{2\epsilon}\int \frac{d^dq}{(2\pi)^d}, 
\end{equation}
where $\Lambda^{2\epsilon}$ refers to the mass dimension of the gauge coupling in $d=4-2\epsilon$ dimensions, which has been absorbed into the master integrals, in this work. Here, this scale relates to the renormalization scale as $\mu^2=4\pi e^{-\gamma}\Lambda^2$.
 
Other master integrals appearing at the four-gluon vertex refer to integrals with three or four propagators such as $\mathcal{C}$ and $\mathcal{D}$:
\begin{widetext}
\begin{align}
    &\mathcal{C}(p_1,p_2,m_0^2,m_1^2,m_2^2)= \!\! \int_q \!\!\ \frac{1}{(q^2+m_0^2)((q+p_1)^2+m_1^2)((q+p_2)^2+m_2^2)},\\
    &\mathcal{D}(p_1,p_2,p_3,m_0^2,m_1^2,m_2^2,m_3^2)= \!\! \int_q \!\!\ \frac{1}{(q^2+m_0^2)((q+p_1)^2+m_1^2)((q+p_2)^2+m_2^2)((q+p_3)^2+m_3^2)}.
\end{align}
In this appendix, we show that in both cases (and in general for any number of propagators, when dealing with collinear configurations), $\mathcal{C}$ and $\mathcal{D}$ can be expressed in terms of $\mathcal{A}$ and $\mathcal{B}$. Therefore, for the class of configurations discussed in this paper, we can give analytical expressions for the four-gluon vertex form factors.

Let us consider first the case of  $\mathcal{C}$ in collinear configurations
\begin{align}
\mathcal{C}= \!\! \int_q \!\!\ \frac{1}{(q^2+m_1^2)((q+\alpha p)^2+m_2^2)((q+\beta p)^2+m_3^2)},
\end{align}
where $\alpha$ and $\beta$ are non-zero real numbers\footnote{If one or both of them are zero the integral can be reduced using $\frac{1}{(q^2+m_1^2)(q^2+m_2^2)}=\frac{1}{m_2^2-m_1^2}\left(\frac{1}{(q^2+m_1^2)}-\frac{1}{(q^2+m_2^2)}\right)$.}. For now on we are going to define $D_1=(q^2+m_1^2)$ to be the first factor in the denominator and also define $D_2=((q+\alpha p)^2+m_2^2)$ and $D_3=((q+\beta p)^2+m_3^2)$.
The integral $\int\frac{d^dq}{(2\pi)^d}\frac{2 p.q}{D_1D_2D_3}$ can be rewritten as
\begin{align}    
\int_q\ \frac{2 p.q}{D_1D_2D_3}&=\frac{1}{\alpha}\int_q\ \left(\frac{D_2-D_1+m_1^2-m_2^2-\alpha^2p^2}{D_1D_2D_3}\right)\nonumber\\
&=\frac{1}{\alpha}\left(\mathcal{B}(\beta p,m_1^2,m_3^2)-\mathcal{B}((\beta-\alpha) p,m_2^2,m_3^2)+(m_1^2-m_2^2-\alpha^2p^2)\mathcal{C}\right).
\end{align}
Analogously, we can express the scalar product in the numerator in terms of the third propagator obtaining:
\begin{align}
\frac{1}{\beta}\left(\mathcal{B}(\alpha p,m_1^2,m_2^2)-\mathcal{B}((\beta-\alpha) p,m_2^2,m_3^2)+(m_1^2-m_3^2-\beta^2p^2)\mathcal{C}\right).
\end{align}
Merging both expressions, we find that
\newpage
\begin{align}
\mathcal{C}=\frac{\alpha\beta}{\alpha(m_1^2-m_3^2-\beta^2p^2)-\beta(m_1^2-m_2^2-\alpha^2p^2)}&\left(\frac{1}{\alpha}\left(\mathcal{B}(\beta p,m_1^2,m_3^2)-\mathcal{B}((\beta-\alpha) p,m_2^2,m_3^2)\right)\right.\nonumber\\
&\left.-\frac{1}{\beta}\left(\mathcal{B}(\alpha p,m_1^2,m_2^2)-\mathcal{B}((\beta-\alpha) p,m_2^2,m_3^2)\right)\right)
\end{align}

Following a similar procedure we can obtain the expression of $\mathcal{D}$ in the collinear kinematics in terms of $\mathcal{C}$ which can be then expressed in terms of $\mathcal{B}$. 

\begin{align}
\mathcal{D}&=\int_q\ \frac{1}{(q^2+m_1^2)((q+\alpha p)^2+m_2^2)((q+\beta p)^2+m_3^2)((q+\gamma p)^2+m_4^2)}\nonumber\\
&=\frac{\alpha\beta}{\alpha(m_1^2-m_3^2-\beta^2p^2)-\beta(m_1^2-m_2^2-\alpha^2p^2)}\left(\frac{1}{\alpha}\left(\mathcal{C}(\beta p,\gamma p,m_1^2,m_3^2,m_4^2)-\mathcal{C}((\beta-\alpha) p,(\gamma-\alpha)p,m_2^2,m_3^2,m_4^2)\right)\right.\nonumber\\
&\left.-\frac{1}{\beta}\left(\mathcal{C}(\alpha p,\gamma p,m_1^2,m_2^2,m_4^2)-\mathcal{C}((\beta-\alpha) p,(\gamma-\alpha)p,m_2^2,m_3^2,m_4^2)\right)\right).
\end{align}

\end{widetext}

		
\end{document}